\documentclass[10pt]{article}
\usepackage{graphicx}

\usepackage{color}

\usepackage[all, warning]{onlyamsmath}
\usepackage{amssymb, amsmath}
\usepackage{dcolumn}
\usepackage{bm}
\usepackage[figuresright]{rotating}
\usepackage{fancyhdr}
\usepackage{enumitem}
\usepackage{siunitx}
\usepackage{indentfirst}
\usepackage{xspace}
\usepackage{color}
\usepackage{epsfig}
\usepackage{grffile}
\usepackage[T1]{fontenc}
\usepackage[utf8]{inputenc}
\usepackage{color}

\newcommand{\etal}{\textsl{et al}.}
\newcommand{\pizero}{\pi^{0}}
\newcommand{\pt}{p_\text{T}}
\newcommand{\sqs}{\sqrt{s}}
\newcommand{\snn}{\sqrt{s_{NN}}}

\newcommand{\ylab}{y_\text{lab}}

\newcommand{\nmf}{R_\textrm{pPb}}

\newcommand{\tev}[1]{\SI{#1}{\tera\electronvolt}}

\def\Title#1{\begin{center} {\Large #1 } \end{center}}
\def\Author#1{\begin{center}{ \sc #1} \end{center}}
\def\Address#1{\begin{center}{ \it #1} \end{center}}

\newcommand{\pubblock}{\rightline{\begin{tabular}{l} Proceedings of the Second
Annual LHCP\\ \pubnumber\\
         \pubdate  \end{tabular}}}

\newenvironment{Abstract}{\begin{quotation} \begin{center} 
             \large ABSTRACT \end{center}\bigskip 
      \begin{center}\begin{large}}{\end{large}\end{center} \end{quotation}}

\newenvironment{Presented}{\begin{quotation} \begin{center} 
             PRESENTED AT\end{center}\bigskip 
      \begin{center}\begin{large}}{\end{large}\end{center} \end{quotation}}

\newcommand\pubnumber{ }

\newcommand\pubdate{\today}

\def\affiliation{
On behalf of the LHCf Collaboration, \\
University of Florence and INFN Firenze,\\
Via Sansone 1, 50019 Sesto Fiorentino (Fi), Italy}

\begin{document}

\large
\begin{titlepage}
\pubblock

\vfill
\Title{Measurements of hadron production in $p$--Pb collisions at LHCf}
\vfill

\Author{Gaku Mitsuka}
\Address{\affiliation}
\vfill
\begin{Abstract}

The transverse momentum distribution for inclusive neutral pions in very forward
rapidity region has been measured with the LHCf detector in $p$--Pb collisions
at nucleon-nucleon center-of-mass energies of $\snn = \tev{5.02}$ at the LHC.
The transverse momentum spectra obtained in $p$--Pb collisions show a strong
suppression of the production of neutral pions relative to the spectra in
$p$--$p$ collisions at $\sqs = \tev{5.02}$. The nuclear modification factor is
about 0.1--0.4, which overall agrees with the predictions of several hadronic
interaction Monte Carlo simulations. Furthermore the recent results on the
inclusive energy spectra of forward neutrons in $p$--$p$ collisions at $\sqs =
\tev{7}$ will be discussed.

\end{Abstract}
\vfill

\begin{Presented}
The Second Annual Conference\\
 on Large Hadron Collider Physics \\
Columbia University, New York, U.S.A \\ 
June 2-7, 2014
\end{Presented}
\vfill
\end{titlepage}
\def\thefootnote{\fnsymbol{footnote}}
\setcounter{footnote}{0}
%

\normalsize 


%
%
\section{Introduction}\label{sec:introduction}

The Large Hadron Collider forward (LHCf) experiment~\cite{LHCfTDR} has been
designed to measure the hadronic production cross sections of neutral particles
emitted in very forward angles, including zero degrees, in proton-proton
($p$--$p$) and proton-lead ($p$--Pb) collisions at the LHC.
The LHCf detectors have the capability for precise measurements of forward
high-energy inclusive-particle-production cross sections of photons, neutrons,
and other neutral mesons and baryons.
The discussions in this proceeding concentrate on (1) the inclusive production
rate for neutral pions ($\pizero$'s) in the rapidity range $-11.0 < \ylab <
-8.9$ as a function of the $\pizero$ transverse momentum, and (2) the inclusive
production rate for neutrons in the pseudo-rapidity range $\eta > 8.81$ as a
function of the neutron energy.

This work is motivated by an application to the understanding of
ultrahigh-energy cosmic ray (UHECR) phenomena, which are sensitive to the
particle productions driven by soft and semi-hard QCD at extremely high energy.
Although UHECR observations have made notable advances in the last few years,
critical parts of the analysis depend on Monte Carlo (MC) simulations of air
shower development that are sensitive to the choice of the hadronic interaction
model.
The fact that the lack of knowledge about forward particle production in
hadronic collisions hinders the interpretation of observations of UHECR was
studied in other documents, for example see Ref.~\cite{Ulrich}.

This proceedings is organized as follows. In Sec.~\ref{sec:detector} the LHCf
detectors are described. The analyses results are then presented in
Sec.~\ref{sec:pi0result} and Sec.~\ref{sec:neuresult}.

%
%
\section{The LHCf detectors}\label{sec:detector}

Two independent LHCf detectors, called Arm1 and Arm2, have been installed in the
instrumentation slots of the target neutral absorbers (TANs)~\cite{TAN} located
$\pm$140\,m from the ATLAS interaction point (IP1) and at zero degree collision
angle.
Charged particles produced at IP1 and directed towards the TAN are swept aside
by the inner beam separation dipole magnet D1 before reaching the TAN.
Therefore only neutral particles produced at IP1 enter the LHCf detector.
At this location the LHCf detectors cover the pseudo-rapidity range from 8.7 to
infinity for zero degree beam crossing angle. With a maximum beam crossing angle
of 140\,$\mu$rad, the pseudo-rapidity range can be extended to 8.4 to infinity.
The structure and performance of the LHCf detectors are explained in
Ref.~\cite{LHCfIJMPA}.

%
%
\section{Transverse momentum spectra of neutral pions in $p$--Pb collisions at
$\snn = \tev{5.02}$}\label{sec:pi0result}

Data used in the analysis in this section were taken in three different runs;
The first run was taken in LHC Fill 3478 and the second and third runs were
taken in LHC Fill 3481. The integrated luminosity of the
data was \SI{0.63}{\nano\barn^{-1}} after correcting for the live time of data
acquisition systems.

\begin{figure*}[htbp]
  \centering
  \includegraphics[height=5cm, keepaspectratio]{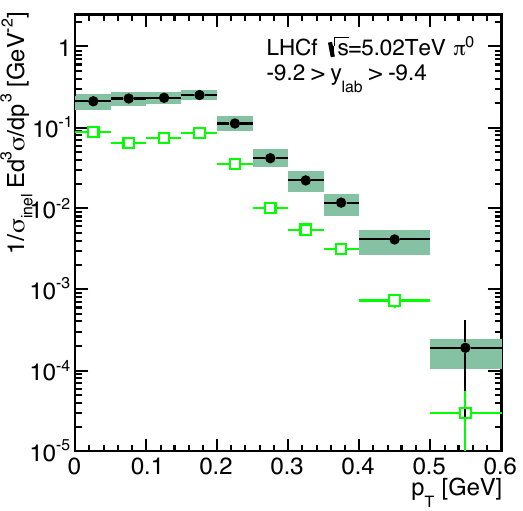}
  \includegraphics[height=5cm, keepaspectratio]{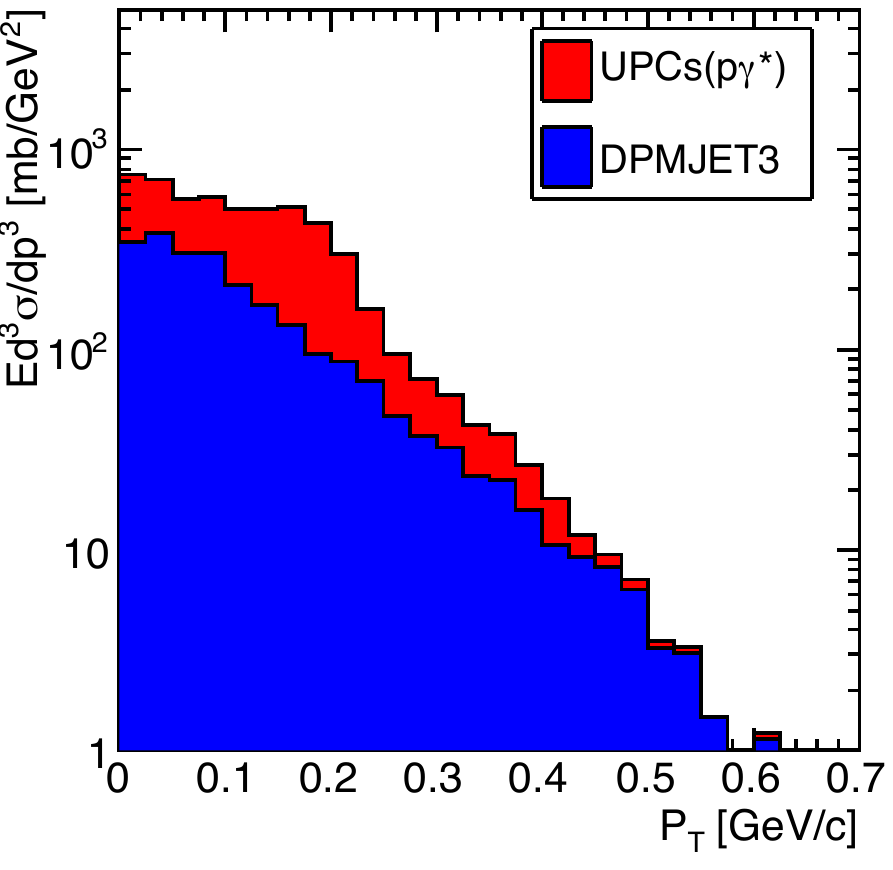}
  \caption{(Left) Filled circles indicate experimental $\pt$ spectra of the LHCf
  detector. The open squares indicate the estimated contribution from UPCs.
  (Right) Estimated $\pt$ spectra of UPC-induced $\pizero$'s (red shaded area)
  and the prediction of a typical hadronic interaction model ({\sc dpmjet 3.04},
  blue shaded area).}
  \label{fig:pt_withupc}
  \centering
\end{figure*}

The $\pt$ spectrum in the rapidity range $-9.4 < \ylab < -9.2$ is presented in
Fig.~\ref{fig:pt_withupc}. The spectra categorized into other five ranges of
rapidity, [-9.0, -8.9], [-9.2, -9.0], [-9.6, -9.4], [-10.0, -9.6], and [-11.0,
-10.0], are found in Ref.~\cite{LHCfpPb}. The inclusive $\pizero$ production
rate is defined as
\begin{equation}
    \frac{1}{\sigma^\text{pPb}_\text{inel}} E\frac{d^{3}\sigma^\text{pPb}}{dp^3}
    =
    \frac{1}{N^\text{pPb}_\text{inel}}
    \frac{d^2 N^\text{pPb}(\pt, y)}{2\pi \pt d\pt dy}.
\end{equation}
\noindent where $\sigma^\text{pPb}_\text{inel}$ is the inelastic cross section
for $p$--Pb collisions at $\snn = \tev{5.02}$ and $E d^{3}\sigma^\text{pPb} /
dp^3$ is the inclusive cross section of $\pizero$ production.
The number of inelastic collisions $N^\text{pPb}_\text{inel}$ in this analysis
is 9.33$\times$10$^{7}$. $d^2N^\text{pPb} (\pt, y)$ is the number of $\pizero$'s
detected in the transverse momentum interval ($d\pt$) and the rapidity interval
($dy$) with all corrections applied.

In the left panel of Fig.~\ref{fig:pt_withupc}, the filled circles represent the
data from the LHCf experiment. The error bars and shaded rectangles indicate the
one-standard-deviation statistical and total systematic uncertainties
respectively. The total systematic uncertainties are given by adding all
uncertainty terms except the one for luminosity in quadrature.
When the impact parameter of beam proton and beam Pb is larger than the
overlapping radii of each particle, so-called ultra-peripheral collisions (UPCs)
can occur.
The contribution from UPCs is presented as open squares in the left panel of
Fig.~\ref{fig:pt_withupc} (normalized to $1/2$ for visibility). The estimation
of the energy and momentum spectra of the UPC-induced $\pizero$'s is discussed
in Ref.~\cite{LHCfpPb} and the comparison with the prediction of a hadronic
interaction model which contains no UPC contribution is shown in the right panel
of Fig.~\ref{fig:pt_withupc}.

\begin{figure*}[htbp]
  \centering
  \includegraphics[width=12cm, keepaspectratio]{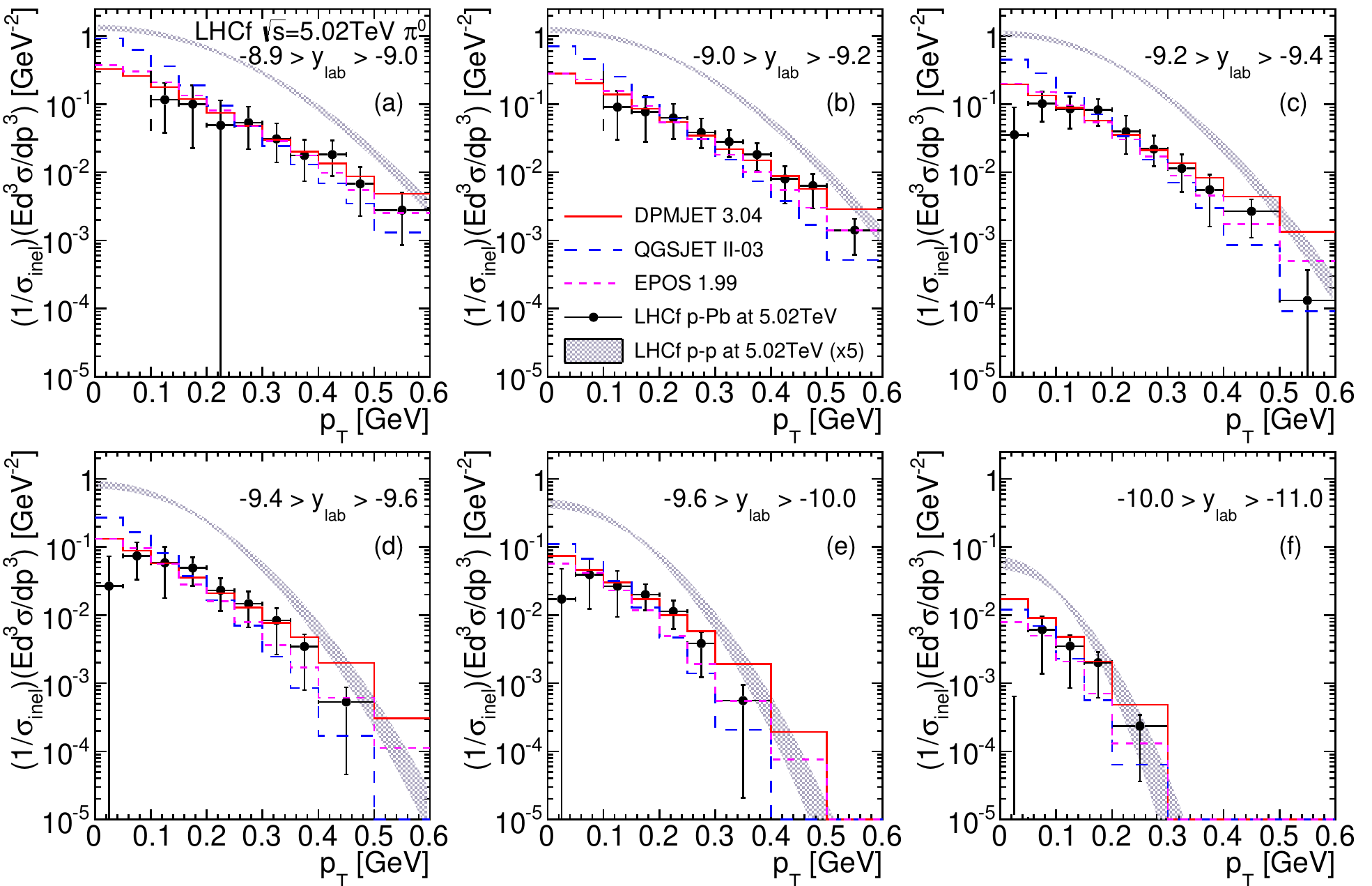}
  \caption{Filled circles indicate experimental $\pt$ spectra measured by LHCf
  after the subtraction of the UPC component. Hadronic interaction models
  predictions and derived spectra for $p$--$p$ collisions at \tev{5.02} are also
  shown.}
  \label{fig:pt_withoutupc}
  \centering
\end{figure*}

Figure~\ref{fig:pt_withoutupc} shows the LHCf data $\pt$ spectra after
subtraction of the UPC component (filled circles). The error bars correspond to
the size of total uncertainties including both statistical and systematic
uncertainties.
The $\pt$ spectra in $p$--Pb collisions at \tev{5.02} predicted by the hadronic
interaction models, {\sc dpmjet 3.04}~\cite{DPM3} (solid line, red), {\sc
qgsjet} II-03~\cite{QGS2} (dashed line, blue), and {\sc epos 1.99}~\cite{EPOS}
(dotted line, magenta), are also shown in the same figure for comparison.
Predictions by the three hadronic interaction models do not include the UPC
component. Among the predictions given by the hadronic interaction models tested
here, {\sc dpmjet 3.04} and {\sc epos 1.99} show a good overall agreement with
the LHCf measurements.

\begin{figure*}[htbp]
  \centering
  \includegraphics[width=12cm,keepaspectratio]{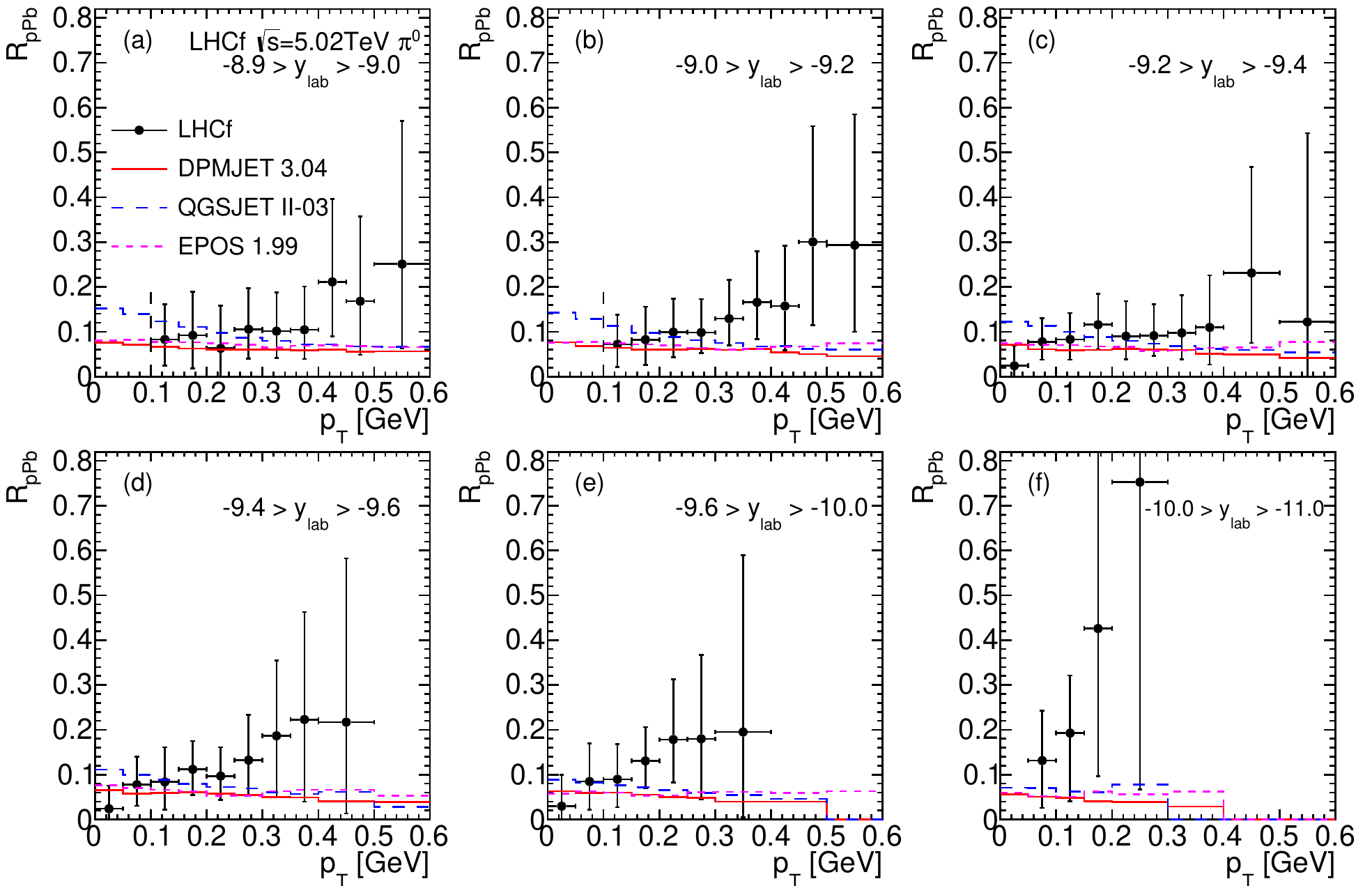}
  \caption{Filled circles indicate the nuclear modification factors obtained by
  the LHCf measurements. Other lines are the predictions by hadronic interaction models
  hadronic interaction models: {\sc dpmjet 3.04} (red solid line), {\sc qgsjet}
  II-03 (blue dashed line), and {\sc epos 1.99} (magenta dotted line).}
  \label{fig:nmf_simple}
  \centering
\end{figure*}

Finally the nuclear modification factor $\nmf$ is derived in
Fig.~\ref{fig:nmf_simple}. The nuclear modification factor is defined as
\begin{equation}
    \nmf \equiv
    \frac{\sigma^\textrm{pp}_\textrm{inel}}
    {\langle N_\textrm{coll} \rangle \sigma^\textrm{pPb}_\textrm{inel}}
    \frac{Ed^3\sigma^\textrm{pPb}/dp^3}{Ed^3\sigma^\textrm{pp}/dp^3},
    \label{eq:nmf}
\end{equation}
\noindent where $E d^{3}\sigma^\text{pPb} / dp^3$ and $E d^{3}\sigma^\text{pp} /
dp^3$ are the inclusive cross sections of $\pizero$ production in $p$--Pb and
$p$--$p$ collisions at \tev{5.02} respectively. The average number of binary
nucleon--nucleon collisions in a $p$--Pb collision is $\langle N_\textrm{coll}
\rangle = 6.9$.

The LHCf measurements, currently having a large uncertainty which increases with
$\pt$ mainly due to systematic uncertainties in $p$--Pb collisions at
$\tev{5.02}$, show a strong suppression with $\nmf$ from 0.1 to 0.4 depending on
$\pt$. All hadronic interaction models predict small values of $\nmf \approx
0.1$, and they show an overall good agreement with the LHCf measurements within
the uncertainty.

%
%
\section{Energy spectra of neutrons in $p$--$p$ collisions at
$\sqs = \tev{7}$}\label{sec:neuresult}

\begin{figure*}[htbp]
  \centering
  \includegraphics[width=15cm,keepaspectratio]{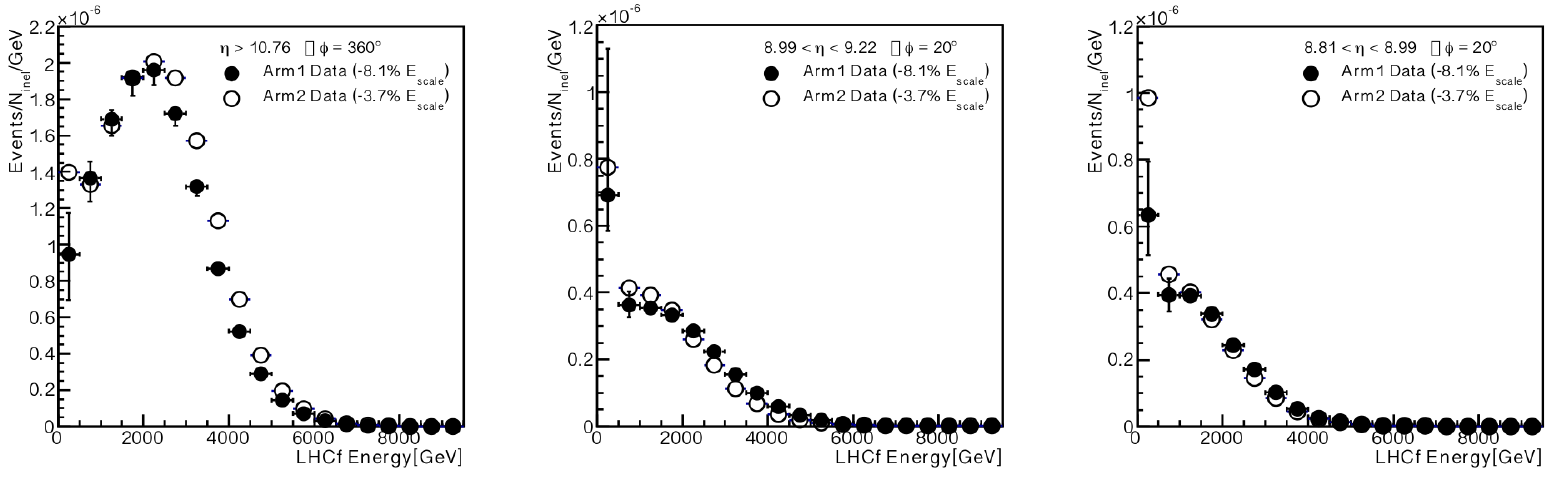}
  \caption{Energy spectra of neutrons measured by the Arm1 (filled circles)
  and the Arm2 detectors (open circles).}
  \label{fig:neutronraw}
  \centering
\end{figure*}

In this section we show the energy spectra of neutrons in $p$--$p$ collisions at
$\sqs = \tev{7}$. The measured energy spectra obtained from the two independent
calorimeters of Arm1 and Arm2 in the pseudo-rapidity $\eta$ ranging from 8.81 to
8.99, from 8.99 to 9.22, and from 10.76 to infinity, are shown in
Fig.~\ref{fig:neutronraw}. The similar characteristic features are found between
the Arm1 and Arm2 measurements in all ranges even before unfolding the spectra.
The slight, but apparent, differences between the two detectors can be
understood by the detector response in each detector.

\begin{figure*}[htbp]
  \centering
  \includegraphics[width=15cm,keepaspectratio]{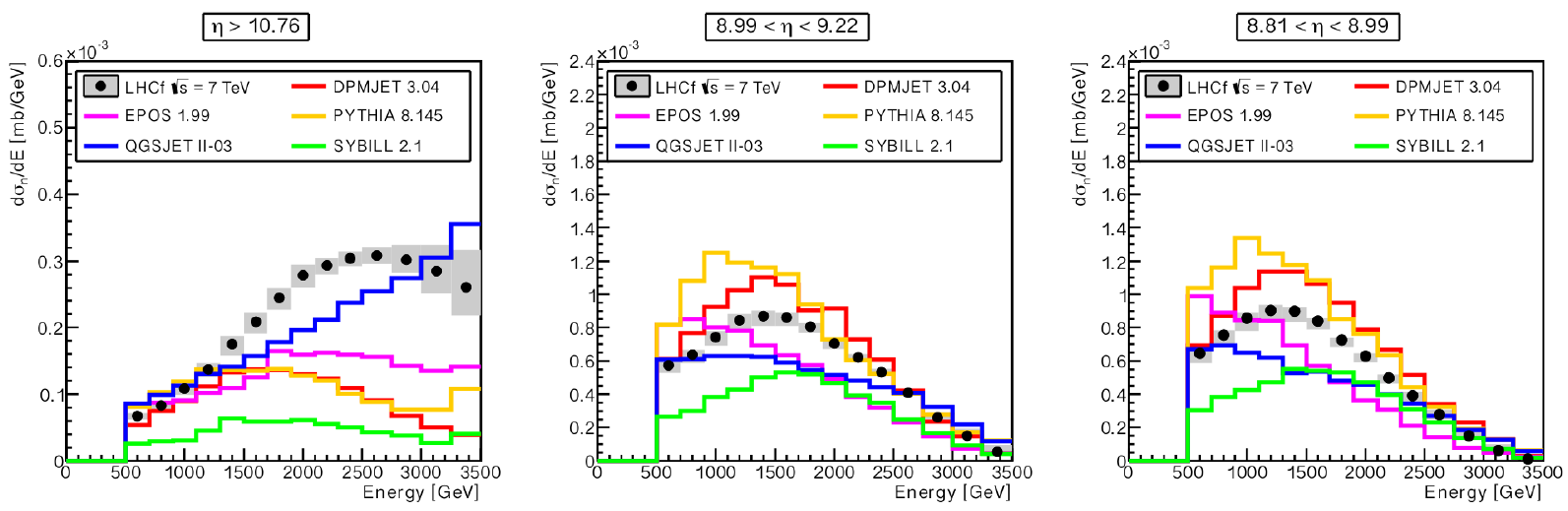}
  \caption{Energy spectra of neutrons obtained by combining with the Arm1 and
  Arm2 measurements (filled circles and shaded areas show the combined results
  and the systematic errors, respectively). Other colored lines are the
  predictions by hadronic interaction models.}
  \label{fig:neutroncomb}
  \centering
\end{figure*}

Figure~\ref{fig:neutroncomb} shows the combined Arm1 and Arm2 spectra with all
corrections and spectra unfolding applied. The experimental results indicate
that the highest neutron production rate occurs at the most forward rapidity
range ($\eta > 10.76$) compared with the hadronic interaction models. {\sc
qgsjet} II-03 predicts a neutron production rate compatible with the
experimental results at the most forward rapidity. On the other hand, {\sc
dpmjet 3.04} shows the best agreement with the experimental results at other two
rapidity ranges.

\end{document}